\pdfoutput = 1

\documentclass[10pt,letterpaper,twocolumn]{article} 
\usepackage{ol2} 
\usepackage{hyperref,cite}

\usepackage{graphicx,ulem}
\usepackage{amsmath,amstext}

\newcommand{\fig}[1]{Fig.~\ref{#1}}

\begin{document}
\twocolumn[
\title{Soliton Coupling Effect: Group-velocity-matching dispersive wave generation and spectral tunneling}
\vspace{-3ex}
\author{Hairun Guo,$^1$ Shaofei Wang,$^2$ Xianglong Zeng,$^{1,2,*}$ Morten Bache$^1$}
\affiliation{
$^1$Ultrafast Nonlinear Optics group, DTU Fotonik, Technical University of Denmark, DK-2800 Kgs. Lyngby, Denmark\\
$^2$Key Laboratory of Special Fiber Optics and Optical Access Networks, Shanghai University, Shanghai 200072, China\\
$^*$Corresponding author: zenglong@shu.edu.cn
}
\begin{abstract}
  Concept of spectral soliton coupling is proposed to explain the group-velocity-matching dispersive wave generation and the soliton spectral tunneling effect. Soliton eigen state is found corresponding to a spectral phase profile under which the soliton phase shifts induced by the dispersion and the nonlinearity are instantaneously counterbalanced. A local but non-eigen soliton pulse will shed off energy and form radiations at other wavelengths, during the coupling to the soliton eigen state. Group-velocity-mathing between the local and generated wave is necessary not only as it produces a coupler-like phase profile which supports soliton coupling but also because it physically supports a soliton phase matching condition, which means the generated wave has a soliton state phase matched to the local exciting pulse. Examples in realistic photonic crystal fiber structures are also presented.

\end{abstract}

\maketitle
]

   Dispersive waves (DWs), also known as optical Cherenkov radiations (OCRs) \cite{akhmediev1995cherenkov}, are generated when temporal solitons are perturbed by higher-order dispersions \cite{wai1986nonlinear}. This phenomenon was experimentally verified soon after its proposal \cite{beaud1987ultrashort,gouveia1988solitons}. DWs play an important role in the attractive octave-spanning supercontinuum generation (SCG) in fiber structures as they dominate the blue-shifted edge of the spectrum while the Raman induced soliton self-frequency shift (SSFS) leads to the red-shifted edge \cite{dudley2010supercontinuum}. The mechanism behind is the phase matching between the DWs and the local soliton wave, which can be shown through a PM topology \cite{stark2011nonlinear}.

   Based on the DW generation, soliton spectral tunneling (SST) effect was proposed \cite{serkin1993soliton} and investigated \cite{tsoy2007theoretical,kibler2007soliton} as a soliton spectral switching phenomenon, in which the DWs are actually generated in another anomalous group velocity dispersion (GVD) region and form a soliton state. Fundamentally, SST is driven by the Raman induced SSFS and requires a potential barrier in the GVD \cite{serkin1993soliton}. A typical GVD barrier is a normal GVD region sandwiched by two anomalous ones, which is usually realizable in photonic crystal fibers (PCFs), with the waveguide dispersion reducing the material dispersion and forming multiple zero-dispersion wavelengths (ZDWs). Moreover, our recent investigation on SST \cite{wang2013optical} showed that along with the PM condition, group velocity (GV) matching is another significant premise for decent soliton switching with high proportion and broad bandwidth.

   In this letter, we point out that the GV-matching DW generation and the SST effect can be understood as a spectral soliton coupling from a local state to its eigen state. Analogous to a spatial waveguide in which eigen modes are supported under a certain phase profile (spatial refractive index profile), for temporal solitons, eigen state is also supported under a spectral phase profile and the soliton phase shifts induced by both the dispersion and the nonlinearity are counterbalanced. DW generation and SST effect are usually evoked by launching a local soliton state but not the eigen state, therefore spectral soliton coupling between the local and eigen states can occur, which shed off the pulse energy and form waves at other wavelegnths. GV-matching condition is essential to tell whether the generated wave is soliton or radiation as it defines the spectral phase profile. If the generated wave is GV-matching to the local wave, the phase profile performs like a coupler structure and the generated wave can form a soliton state, while with GV mismatched, the phase profile is a leaking structure and the generated wave performs as leakage from the local wave, namely forming radiation.

\begin{figure}[hbt]
  \centering{
  \includegraphics[width = 0.5 \linewidth]{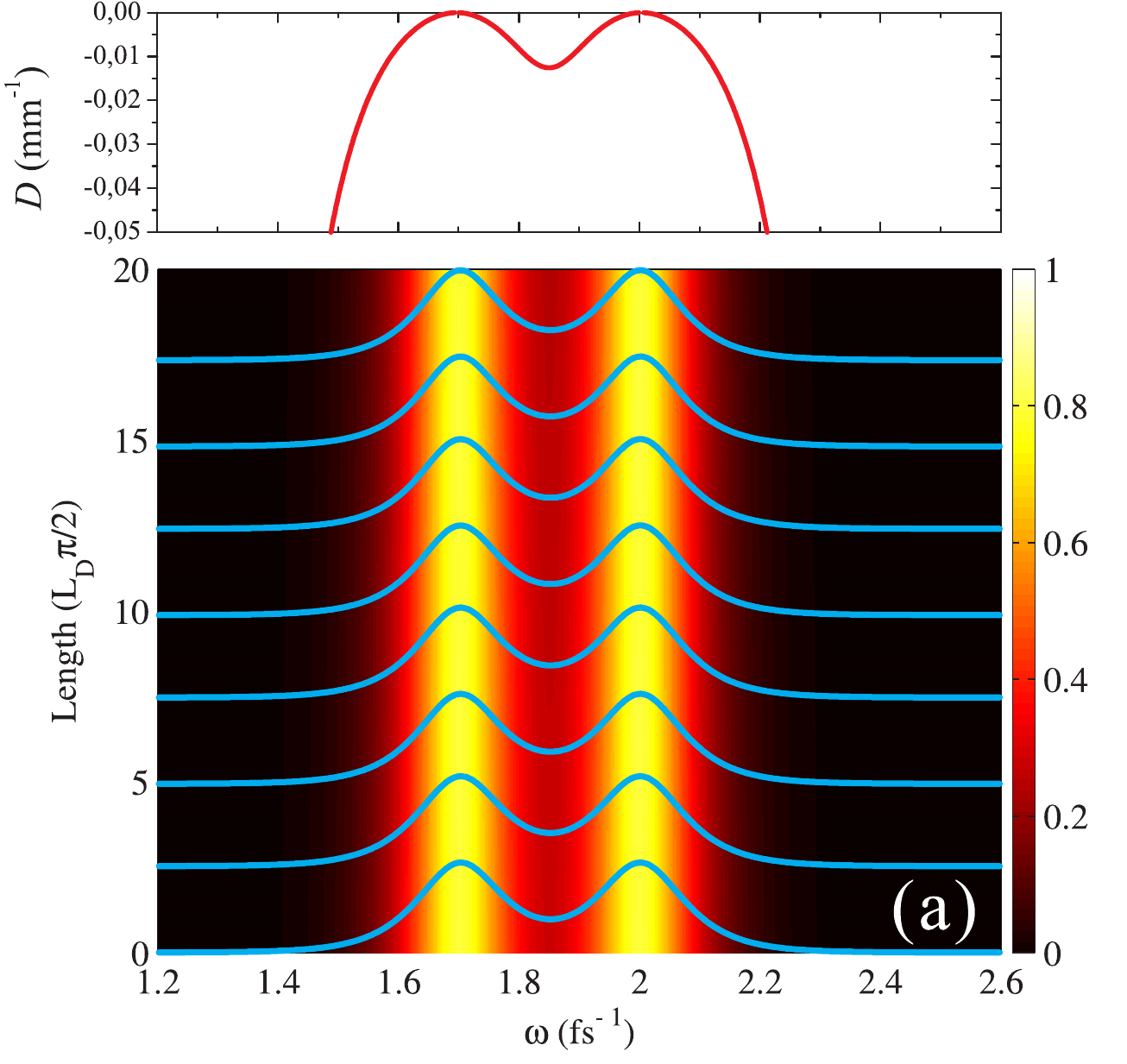}%
  \includegraphics[width = 0.5 \linewidth]{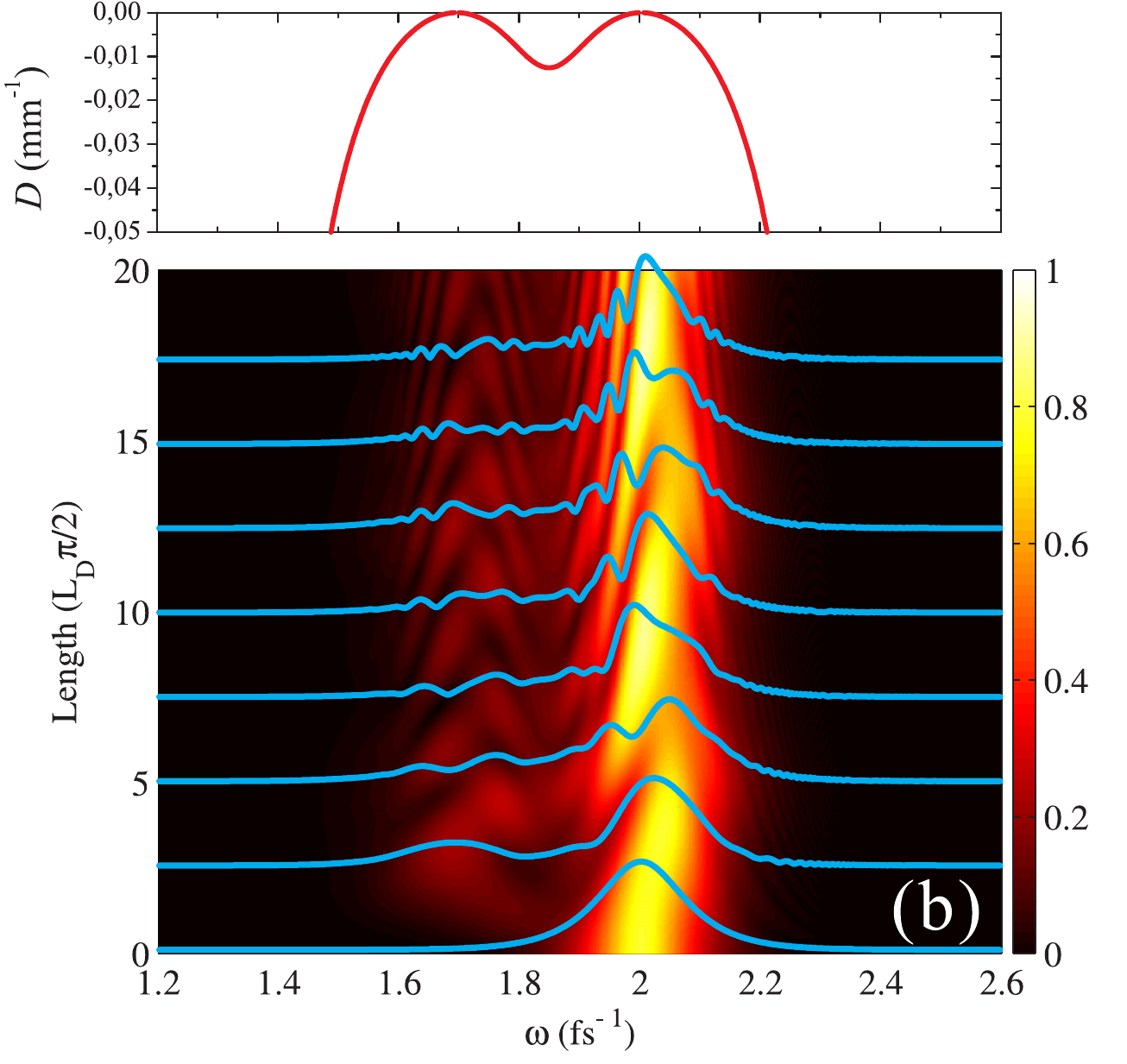}
  \includegraphics[width = 0.5 \linewidth]{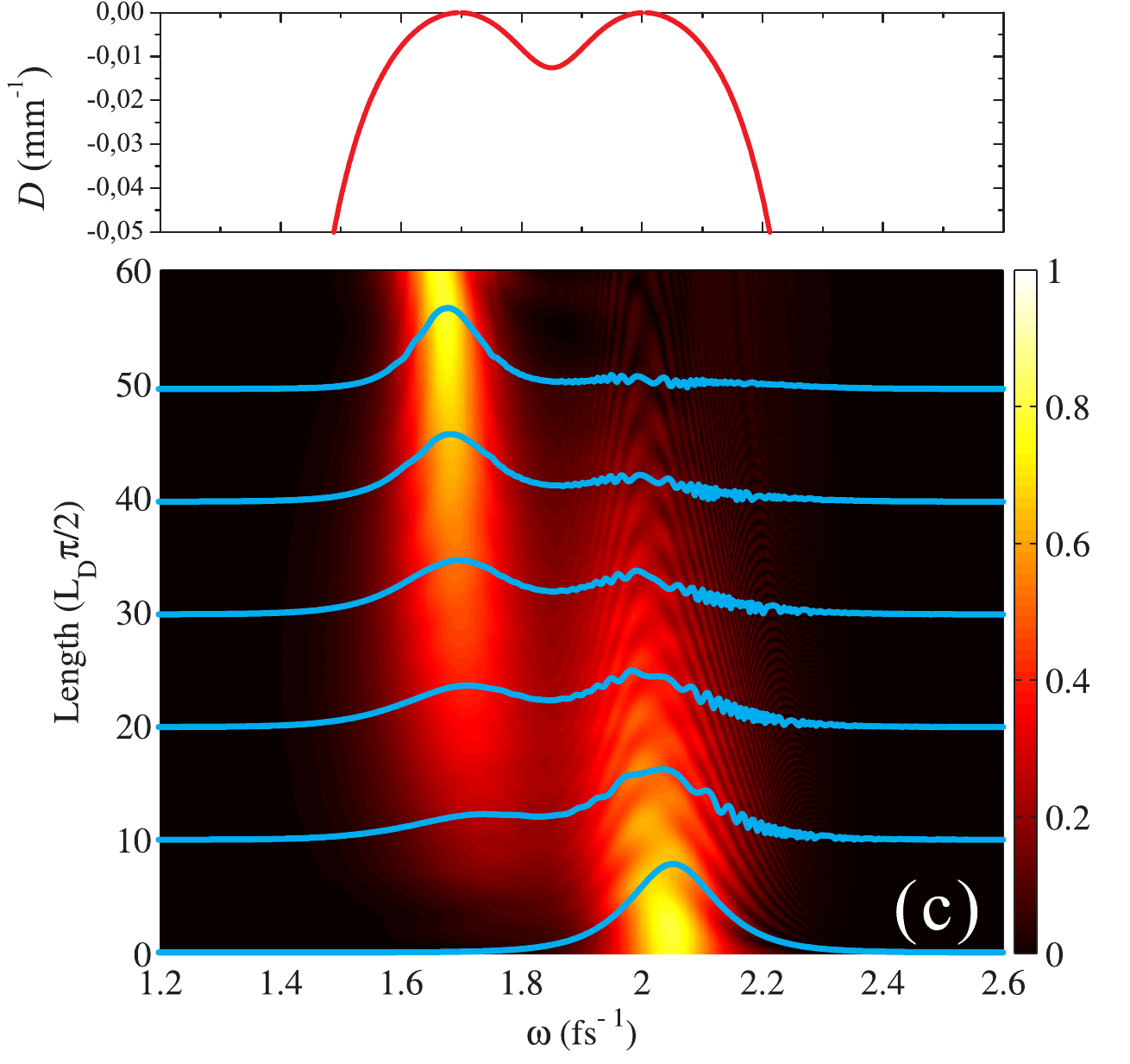}%
  \includegraphics[width = 0.5 \linewidth]{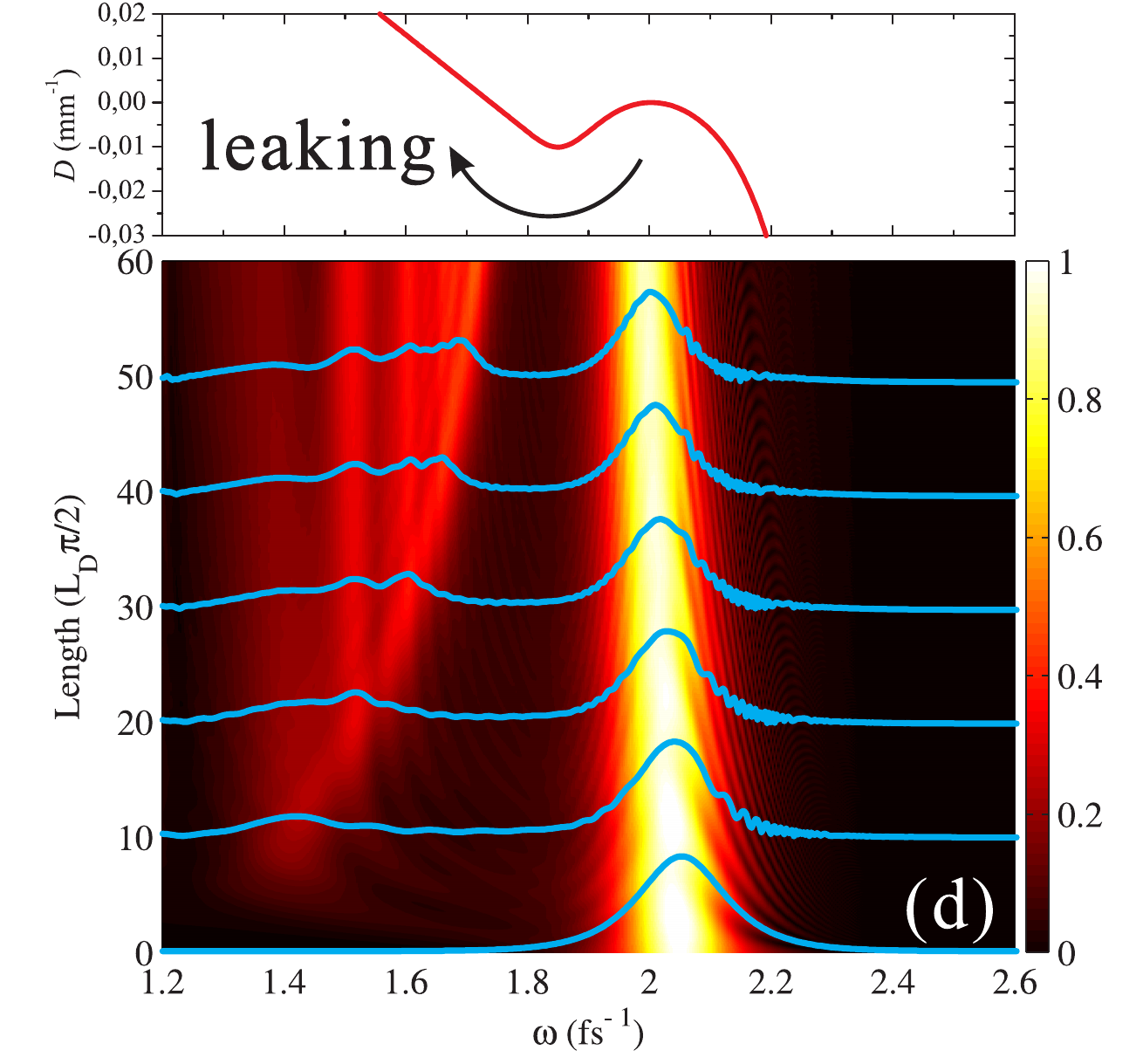}
  \includegraphics[width = 0.5 \linewidth]{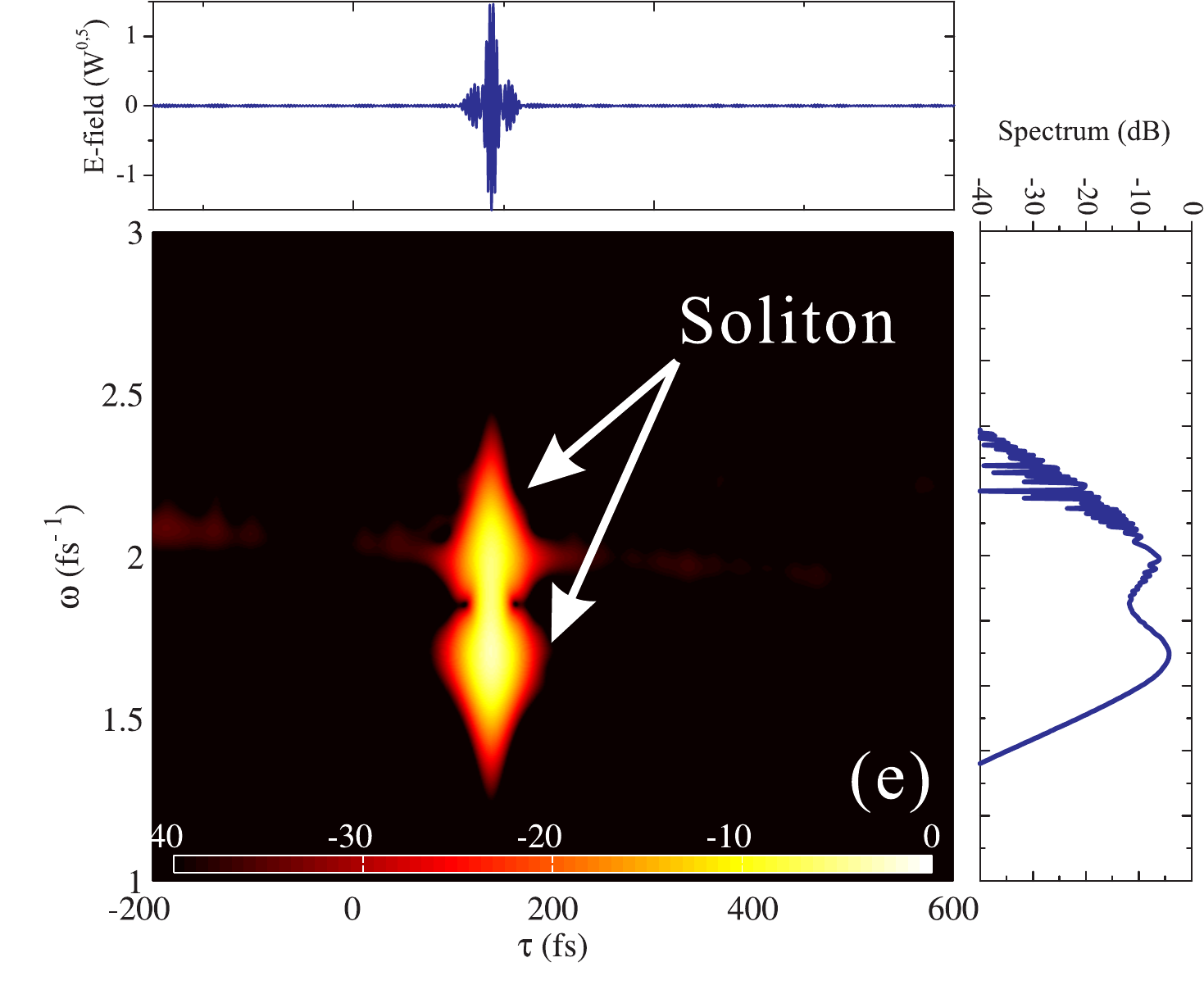}%
  \includegraphics[width = 0.5 \linewidth]{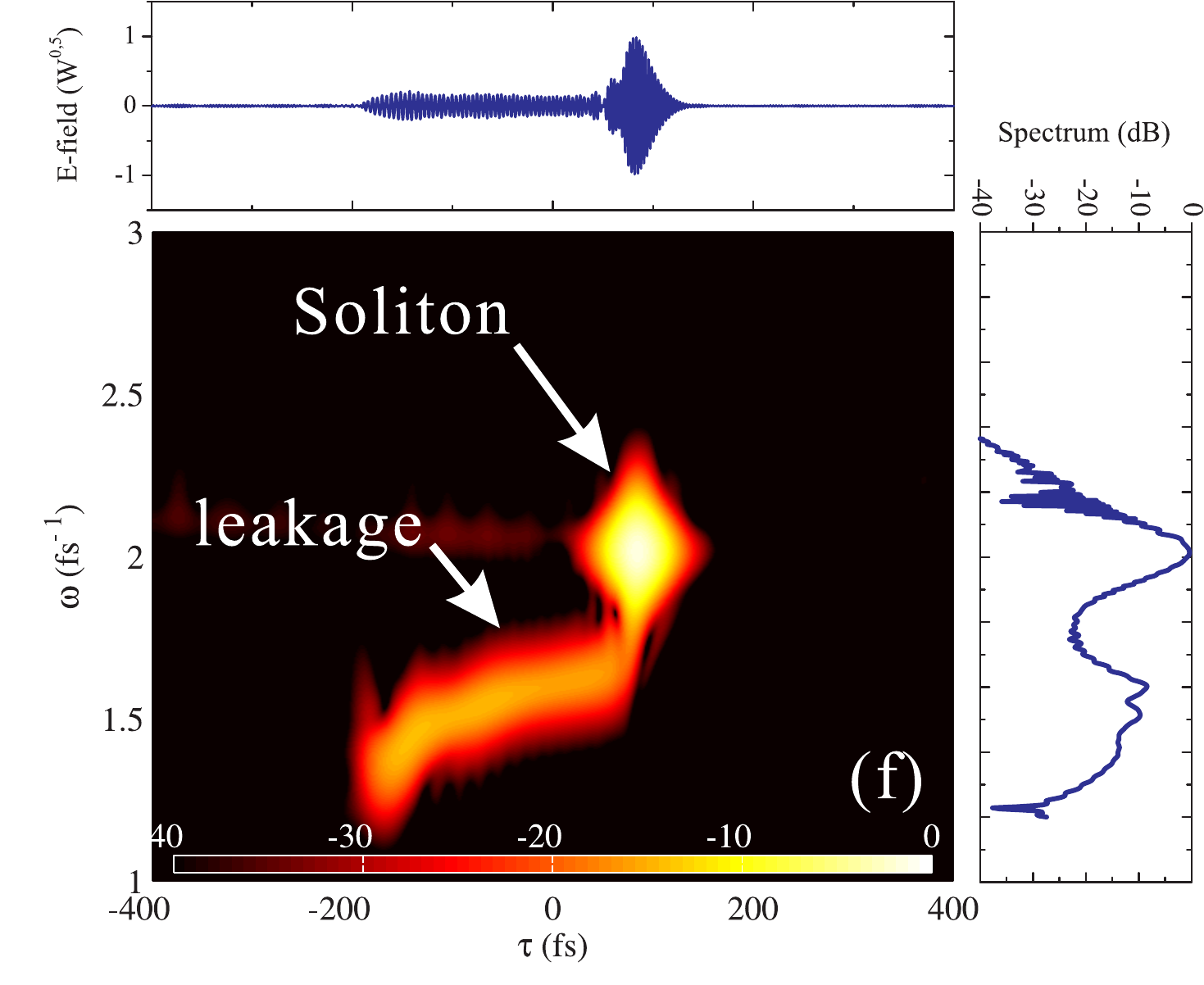}
  }
  \caption{Soliton eigen state, soliton coupling and soliton leaking shown through simulating the NWEF. The nonlinear coefficient ${\gamma = 10 {\rm (km\cdot W)^{-1}}}$. (a) soliton eigen state of a coupler-like phase profile, each soliton peak has power 0.4kW; (b) soliton coupling from the local state to the eigen state. The local soliton has hyperbolic secant shape, the soliton order is set to unity, FWHM = 17.63 fs; (c) SST effect with extra 20\% material Raman effects. The Raman spectral response is chosen from Ref.\cite{roy2009effects}; (d) soliton leaking in a leaking phase profile; (e) and (f) XFROG pattern of (c) and (d) at half propagation distance, with the corresponding spectra and profils of the electric field; Gate function in the XFROG pattern has a hyperbolic secant shape with FWHM = 17.63 fs.}
  \label{Fig-1}
\end{figure}

   We begin with the demonstration of the soliton eigen state. The model behind is the nonlinear wave equation in frequency domain (NWEF) \cite{guo2013generalized,guo2013nonlinear} with the dispersion and the cubic nonlinearity included, i.e. ${{\textstyle{\partial  \over {\partial z}}}\tilde E + i\beta \tilde E + i\gamma F[{\left| E \right|^2}E] = 0}$ , where ${\tilde E}$ is the electric field written in frequency domain, ${\beta}$ refers to the physical phase, ${\gamma}$ is the nonlinear coefficient and ${F}$ indicates the Fourier transform. Pulse self-steepening is automatically included in such a spectral equation. The phase profile is defined with the initial phase ${\beta _0}$ and the global GV ${v_{g,0}}$ eliminated from the physical phase, i.e. ${\beta _{\rm eig}(\omega ) = \beta (\omega ) - {\beta _0} - \omega  \cdot v_{g,0}^{ - 1}}$. Shown in \fig{Fig-1}(a), a coupler-like phase profile is demonstrated to support a soliton eigen state which consists of two spectral peaks corresponding to the coupler channels.

   If only a local soliton is launched into the above phase profile, the soliton spectral coupling is observed as shown in \fig{Fig-1}(b). The local wave shed off energy into the adjacent channel and form a wave. Meanwhile, the local soliton will experience a spectral recoil effect \cite{chang2011fiber} away from the coupling. The energy coupled out will also be back since they have the GV-matching condition, and the local soliton will be pulled back and then evokes the coupling over again. The generated wave (used to be understood as DW) here is actually a soliton wave which travels together with the local soliton, it is fed by the local soliton periodically but unfortunately the energy is low due to the recoil effect.

   Including material Raman effects, SST effect is observed as shown in \fig{Fig-1}(c). Since the Raman response induces SSFS which is always red-shifted and could be against the recoil effect, the local soliton can be kept at its position and the soliton coupling continuously occurs until the local wave are fully coupled into the adjacent channel to form a new soliton wave. Afterwards, the new soliton will also experience the Raman induced SSFS, which means the coupling backwards are greatly suppressed since the soliton will be red-shifted away from the coupling. The XFROG pattern of the SST also shows that the generated soliton has the GV matching with the local one, see \fig{Fig-1}(e).

   However, in a leaking structure of the phase profile where the generated wave has GV mismatched to the launched soliton, SST will not occur, shown in \fig{Fig-1}(d). In the XFROG pattern, waves are gradually radiated as the leakage from the local soliton, leaving a long tail in the pattern, see \fig{Fig-1}(f).

   In fact, the GV-matching condition helps turning the DW PM condition into a soliton PM condition. The DW PM condition can be written as:

   \begin{equation}
   \beta ({\omega _r}) = {\beta _{sol,{\omega _s}}}({\omega _r})
   \label{Eq-OCR-PM}
   \end{equation}
   where ${{\beta _{sol,{\omega _s}}}(\omega ) = \beta ({\omega _s}) + (\omega  - {\omega _s})v_{g,s}^{ - 1} + {q_s}}$ represents the non-dispersive soliton phase with spectrum centered at ${\omega _s}$,  ${q_s}$ is the soliton wave number and, for fundamental solitons, its contribution is minimal \cite {bache2010optical}. Therefore, the above equation can be expanded as:

   \begin{equation}
   \beta ({\omega _r}) + (\omega  - {\omega _r}) \cdot v_{g,s}^{ - 1} = \beta ({\omega _s}) + (\omega  - {\omega _s})v_{g,s}^{ - 1}
   \label{Eq-OCR-PM-expnd}
   \end{equation}

   With GV-matching condition, ${v_{g,r} = v_{g,s}}$, the above equation finally becomes ${{\beta _{sol,{\omega _s}}}(\omega ) = {\beta _{sol,{\omega _r}}}(\omega )}$, implying the phase matching within the whole frequency domain between two solitons located at ${\omega _s}$ and ${\omega _r}$.

\begin{figure}[htb]
  \centering{
  \includegraphics[width = 1 \linewidth]{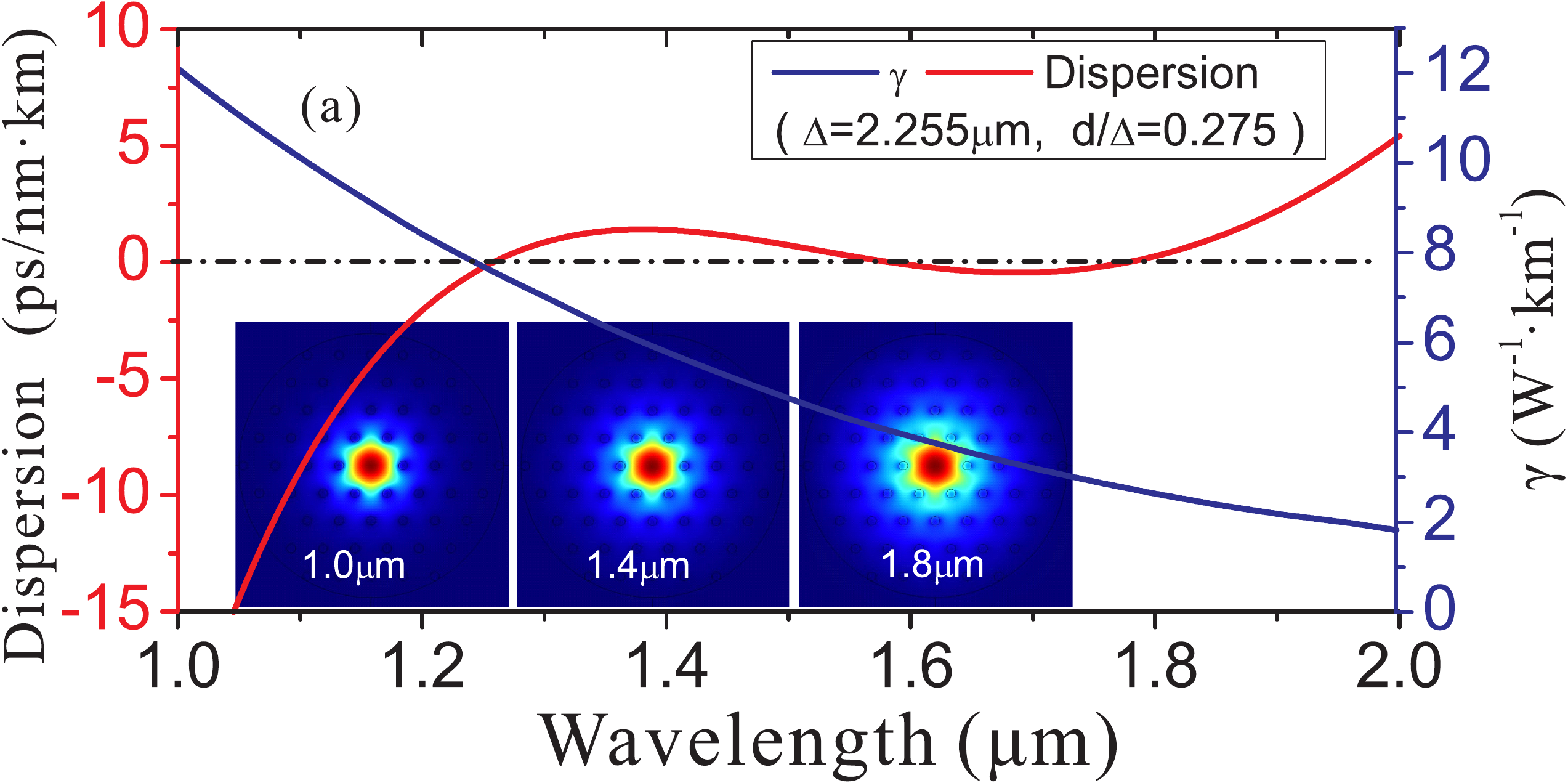}
  }
  \caption{Solid-core index guiding PCF's dispersion and effective nonlinear coefficient $\gamma$ versus wavelength, estimated by means of Comsol software. Insets are mode field distributions at different wavelengths.}
  \label{Fig-2}
\end{figure}

   In practice, PCFs with designed pitch ${\Delta }$ and hole ${d}$ sizes can achieve flexible dispersion profiles with multiple controlled ZDWs. For example, in \fig{Fig-2}, a solid-core index-guiding PCF with a triangular air-hole partten in the cladding has form a concave-like dispersion profile with 3 ZDWs. Such dispersion profile is actually produced by the mode coupling between the core and the air-hole cladding around a resonant wavelength. It is noticed that at long wavelengths, the mode field distribution is expanded into the cladding and therefore the nonlinear coefficient is reduced as the effective mode area is increased.

\begin{figure}[htb]
  \centering{
  \includegraphics[width = 1 \linewidth]{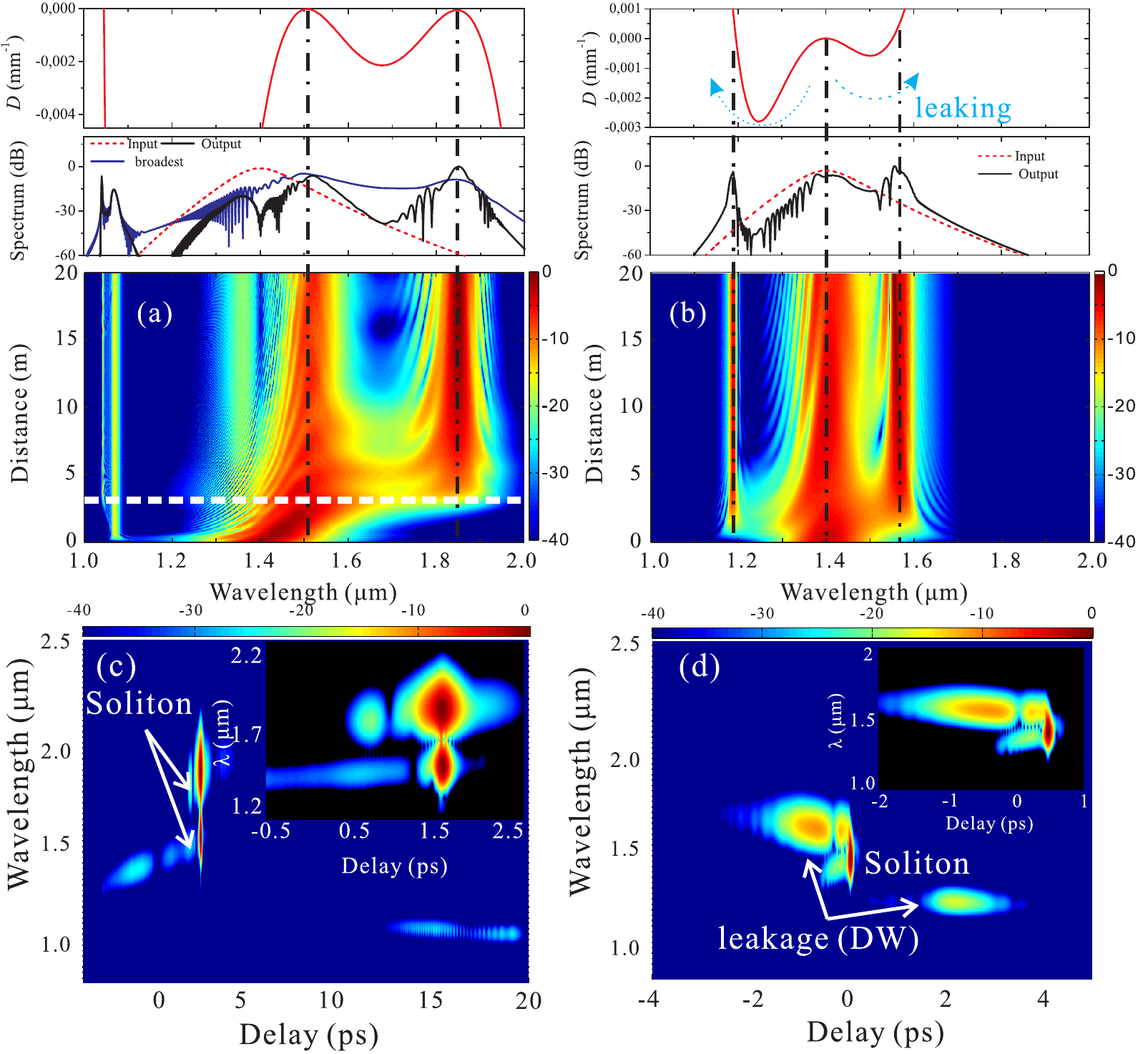}
  }
  \caption{NWEF simulations on the spectral evolutions of a 25 fs (FWHM) local fundamental soliton centered at ${1.4 {\rm \mu m}}$ under the coupler-like and leaking phase profiles: (a) coupler-like structure with ${\Delta = 2.255 {\rm \mu m}}$ and (b) leaking structuere with ${\Delta = 2.235 {\rm \mu m}}$. The input, output spectra of both cases and broadest spectrum during the SST (position marked by white dashed line) are shown. The input pulse has a peak power 380W, the material Kerr nonlinearity is ${2.6 \times 10^{-20} {\rm m^2/W}}$, Raman fraction is 24.5\%. (c) and (d) XFROG pattern of (a) and (b) at a propagation length of 20m. The Gate function in XFROG pattern also has a FWHM = 25 fs.}
  \label{Fig-3}
\end{figure}

   The phase profile of such a PCF structure is plotted in \fig{Fig-3}(a), by eliminating the ${\beta _0}$ and ${v_{g,0}}$ around ${1.5 {\rm \mu m}}$, showing a coupler-like profile. Under such a profile, a local fundamental soliton centered at ${1.4 {\rm \mu m}}$ is launched, which is little away from the coupling. Hence, during the propagation, the local soliton will first experience the Raman induced SSFS, red-shifting towards the coupling position. Then, strong soliton coupling occurs with most of the energy transferred into the adjacent channel, centered at ${1.85 {\rm \mu m}}$. Actually, the choose of the reference-like ${\beta _0}$ and ${v_{g,0}}$ is directly linked to the PM condition. A symmetry coupler-like phase profile, with the same peak height, is corresponding to the PM between the local and nonlocal solitons, under which the soliton coupling is strongest. The XFROG pattern in \fig{Fig-3}(c) again proves the GV matching between the two solitons. The proportion of the soliton energy transfer is around 60\%, lower than the ideal SST because the nonlinear coefficient ${\gamma }$ has a reduction in long wavelengths.

   With a different pitch size, the phase profile turns to a leaking structure. Therefore, the soliton leaking occurs instead of coupling, see \fig{Fig-3}(b). Although from the spectral evolution, the nonlocal wave generated are still decent, they are actually not soliton waves but just leakage-like DWs, proved by the XFROG pattern in \fig{Fig-3}(d).

   As a conclusion, we propose the concept of soliton coupling from a local state to the eigen state, which helps us properly explain the DW generation and SST effect. The basis of the soliton coupling is the soliton eigen state under a spectral phase profile, in which the soliton phase shifts induced by both the dispersion and the nonlinearity are counterbalanced. Under a coupler-like phase profile, soliton eigen state is demonstrated to have a two-peak profile, and a local but non-eigen soliton is therefore expected to excite the soliton coupling. Sepctral recoil effect is observed during the coupling, which always pushes the pulse away from the coupling position and suppresses the coupling. With material Raman effects, red-shifted SSFS is produced and could be against the recoil effect, which therefore helps keep the pulse at its position and causes continuously soliton coupling, known as SST effect. The GV-matching condition is crucial not only as it produces such a coupler-like phase profile, but also because it physically turns a DW PM condition into a soliton PM condition. With GV mismatched between the local pulse and the generated wave, the phase profile is supposed to be a leaking structure, under which the generated wave performs as leakage from the local pulse. Such soliton coupling and soliton leaking effects in realistic PCF structures are presented, in which SST with a spectral span over 300 nm and 60\% energy transferring is numerically demonstrated.

   This work was supported in part by National Natural Science Foundation of China (60978004,11274224) and Shanghai Shuguang Program (10SG38). Xianglong Zeng acknowledges Marie Curie International Incoming Fellowship (PIIF-GA-2009-253289). Morten Bache acknowledges the Danish Council for Independent Research 11-106702.

\begin{thebibliography}{10}
\newcommand{\enquote}[1]{``#1''}

\bibitem{akhmediev1995cherenkov}
N.~Akhmediev and M.~Karlsson, \enquote{Cherenkov radiation emitted
  by solitons in optical fibers,} Phys. Rev. A \textbf{51}, 2602--2607
  (1995).

\bibitem{wai1986nonlinear}
P.~Wai, C.~R. Menyuk, Y.~Lee, and H.~Chen, \enquote{Nonlinear pulse propagation
  in the neighborhood of the zero-dispersion wavelength of monomode optical
  fibers,} Opt. Lett. \textbf{11}, 464--466 (1986).

\bibitem{beaud1987ultrashort}
P.~Beaud, W.~Hodel, B.~Zysset, and H.~Weber, \enquote{Ultrashort pulse
  propagation, pulse breakup, and fundamental soliton formation in a
  single-mode optical fiber,} IEEE J. Quant. Electron. \textbf{23},
  1938--1946 (1987).

\bibitem{gouveia1988solitons}
A.~Gouveia-Neto, M.~E. Faldon, and J.~Taylor, \enquote{Solitons in the region
  of the minimum group-velocity dispersion of single-mode optical fibers,}
  Opt. Lett. \textbf{13}, 770--772 (1988).

\bibitem{dudley2010supercontinuum}
J.~M. Dudley and J.~R. Taylor, \enquote{Supercontinuum generation in optical
  fibers} (Cambridge University Press, 2010).

\bibitem{stark2011nonlinear}
S.~Stark, F.~Biancalana, A.~Podlipensky, and P.~S.~J. Russell,
  \enquote{Nonlinear wavelength conversion in photonic crystal fibers with
  three zero-dispersion points,} Phys. Rev. A \textbf{83}, 023808 (2011).

\bibitem{serkin1993soliton}
V.~Serkin, V.~Vysloukh, and J.~Taylor, \enquote{Soliton spectral tunnelling
  effect,} Electron. Lett. \textbf{29}, 12--13 (1993).

\bibitem{tsoy2007theoretical}
E.~N. Tsoy and C.~M. de~Sterke, \enquote{Theoretical analysis of the
  self-frequency shift near zero-dispersion points: Soliton spectral
  tunneling,} Phys. Rev. A \textbf{76}, 043804 (2007).

\bibitem{kibler2007soliton}
B.~Kibler, P.-A. Lacourt, F.~Courvoisier, and J.~Dudley, \enquote{Soliton
  spectral tunnelling in photonic crystal fibre with sub-wavelength core
  defect,} Electron. Lett. \textbf{43}, 967--968 (2007).

\bibitem{wang2013optical}
S.~Wang, J.~Hu, H.~Guo, and X.~Zeng, \enquote{Optical cherenkov radiation in an
  ${\rm As_2S_2}$ slot waveguide with four zero-dispersion
  wavelengths,} Opt. Express \textbf{21}, 3067--3072 (2013).

\bibitem{guo2013generalized}
H.~Guo, X.~Zeng, and M.~Bache, \enquote{Generalized nonlinear wave equation in
  frequency domain,} arXiv preprint arXiv:1301.1473  (2013).

\bibitem{guo2013nonlinear}
H.~Guo, X.~Zeng, B.~Zhou, and M.~Bache, \enquote{Nonlinear wave equation in
  frequency domain: accurate modeling of ultrafast interaction in anisotropic
  nonlinear media,} J. Opt. Soc. Am. B \textbf{30}, 494--504 (2013).

\bibitem{chang2011fiber}
G.~Chang, L.~Chen, and K. Franz X, \enquote{Fiber-optic Cherenkov radiation in the few-cycle regime,} Opt. Express \textbf{19} 6635--6647 (2011).

\bibitem{roy2009effects}
S.~Roy, S.~K. Bhadra, and G.~P. Agrawal, \enquote{Effects of higher-order
  dispersion on resonant dispersive waves emitted by solitons,} Opt. Lett.
  \textbf{34}, 2072--2074 (2009).

\bibitem{bache2010optical}
M.~Bache, O.~Bang, B.~Zhou, J.~Moses, and F.~Wise, \enquote{Optical cherenkov
  radiation in ultrafast cascaded second-harmonic generation,} Phys. Rev.
  A \textbf{82}, 063806 (2010).


\end{thebibliography}

\clearpage

\end{document}